\documentclass{article}
\usepackage{amsmath}
\usepackage{amsfonts}
\usepackage{amssymb}
\usepackage{epsfig}
\usepackage{url}
\usepackage{graphicx}
\usepackage{dcolumn}
\usepackage{bm}

\begin{document}
\title{Optimal Data-Based Binning for Histograms}
\author{Kevin H. Knuth\\ Departments of Physics and Informatics\\
University at Albany (SUNY)\\ Albany NY 12222, USA}

\date{\today}
\maketitle

\begin{abstract}
Histograms are convenient non-parametric density
estimators, which continue to be used ubiquitously.  Summary
quantities estimated from histogram-based probability density
models depend on the choice of the number of bins. We introduce a
straightforward data-based method of determining the optimal
number of bins in a uniform bin-width histogram. By assigning a multinomial likelihood and a non-informative prior, we derive the posterior probability for the
number of bins in a piecewise-constant density model given the
data. In addition, we estimate the mean and standard deviations of the resulting bin heights, examine the effects of
small sample sizes and digitized data, and demonstrate the
application to multi-dimensional histograms.
\end{abstract}



\section{\label{sec:level1}Introduction}

Histograms are used extensively as nonparametric density
estimators both to visualize data and to obtain summary
quantities, such as the entropy, of the underlying density.
However in practice, the values of such summary quantities depend
on the number of bins chosen for the histogram, which given the
range of the data dictates the bin width. The idea is to choose a
number of bins sufficiently large to capture the major features in
the data while ignoring fine details due to `random sampling
fluctuations'. Several rules of thumb exist for determining the
number of bins, such as the belief that between 5-20 bins is
usually adequate (for example, \texttt{Matlab} uses 10 bins as a
default). Scott \cite{Scott:1979, Scott:1992} and
Freedman and Diaconis \cite{Freedman&Diaconis:1981} derived
formulas for the optimal bin width by minimizing the integrated
mean squared error of the histogram model $h(x)$ of the true
underlying density $f(x)$,
\begin{equation}
L(h(x), f(x)) = \int{dx \bigl(h(x)-f(x)\bigr)^2}.
\end{equation}
For $N$ data points, the optimal bin width $v$ goes as $\alpha
N^{-1/3}$, where $\alpha$ is a constant that depends on the form
of the underlying distribution. Assuming that the data are
normally distributed with a sample variance $s$ gives $\alpha =
3.49s$ \cite{Scott:1979, Scott:1992}, and
\begin{equation}
v_{\mathrm{scott}} = 3.49s N^{-1/3}.
\end{equation}
Given a fixed range $R$ for the data, the number of bins $M$ then
goes as
\begin{equation}
M_{\mathrm{scott}} = \lceil \frac{R}{3.49 s}
N^{1/3}\rceil.\label{eq:scott}
\end{equation}
Freedman and Diaconis report similar results, however they suggest
choosing $\alpha$ to be twice the interquartile range of the data.
While these appear to be useful estimates for unimodal densities
similar to a Gaussian distribution, they are known to be
suboptimal for multimodal densities. This is because they were
derived by assuming particular characteristics of the underlying
density. In particular, the result obtained by Freedman and Diaconis is not
valid for some densities, such as the uniform density, since it
derives from the assumption that the density $f$ satisfies
$\int{f'^2>0}$.

Another approach by Stone \cite{Stone:1984} relies on minimizing $L(h,f) - \int{f^2}$ to obtain a rule
where one chooses the bin width $v$ to minimize
\begin{equation}\label{eq:stone}
K(v,M) = \frac{1}{v}\biggl( \frac{2}{N-1} - \frac{N+1}{N-1}
\sum_{m=1}^{M}{\pi_i^2} \biggr)
\end{equation}
where $M$ is the number of bins and $\pi_i$ are the bin
probabilities.  Rudemo obtains a similar rule
by applying cross-validation techniques with a Kullback-Leibler
risk function \cite{Rudemo:1982}.

We approach this problem from a different perspective. Since the
underlying density is not known, it is not reasonable to use an
optimization criterion that relies on the error between our
density model and an unknown true density. Instead, we consider the
histogram to be a piecewise-constant model of the underlying
probability density. Using Bayesian probability theory we derive a
straightforward algorithm that computes the posterior probability
of the number of bins for a given data set. Within this framework defined by our likelihood and prior probability assignments,  one can objectively select an optimal piecewise-constant model describing
the density function from which the data were sampled.

It should be emphasized that this paper considers equal bin width piecewise-constant density models where one possesses little to no prior information about the underlying density from which the data were sampled.  In many applications, variable bin width models \cite{Wegman:1975, Denison:2002, Scargle:2005, Endres&Foldiak:2005, Hutter:2007} may be more efficient or appropriate, and certainly if one possesses prior information about the underlying density, a more appropriate model should be considered.

\section{The Piecewise-Constant Density Model}
We are given a dataset consisting of $N$ data values that were sampled from an unknown probability density function.  The sampled data values are assumed to be known precisely so that there is no additional measurement uncertainty associated with each datum.  We begin by considering the histogram as a piecewise-constant
model of the probability density function from which $N$ data
points were sampled. This model has $M$ bins with each bin having
equal width $v = v_k$, where $k$ is used to index the bins.  Together they encompass an entire range of data values $V = Mv$. Note that for a one-dimensional histogram, $v_k$ is the width of the $k^{th}$ bin.
In the case of a multi-dimensional histogram, this will be a
multi-dimensional volume. Each bin has a ``height'' $h_k$, which
is the constant probability density over the region of the bin.
Integrating this constant probability density $h_k$ over the width
of the bin $v_k$ leads to a probability mass of $\pi_k = h_k
v_k$ for the bin. This results in the following piecewise-constant
model $h(x)$ of the unknown probability density function $f(x)$
\begin{equation}
h(x) = \sum_{k = 1}^{M}{h_k~\Pi(x_{k-1}, x, x_k)},
\end{equation}
where $h_k$ is the probability density of the $k^{th}$ bin with
edges defined by $x_{k-1}$ and $x_k$, and $\Pi(x_{k-1}, x, x_k)$
is the boxcar function where
\begin{equation}
\Pi(x_a, x, x_b) =
    \left\{ \begin{array}{rl}
       0 & \mbox{if}~~x < x_a\\
       1 & \mbox{if}~~x_a \leq x < x_b\\
       0 & \mbox{if}~~x_b \leq x \end{array} \right.
\end{equation}
This density model can be re-written in terms of the bin
probabilities $\pi_k$ as
\begin{equation}
h(x) = \frac{M}{V} \sum_{k = 1}^{M}{\pi_k~\Pi(x_{k-1}, x, x_k)}.
\end{equation}
It is important to keep in mind that $h(x)$ is not a histogram, but rather it is a piecewise-constant probability density function.  The bin heights $h_k$ represent the probability density assigned to the $k^{th}$ bin, and the parameters $\pi_k$ represents the probability mass of the $k^{th}$ bin.

Given $M$ bins and the normalization condition that the integral
of the probability density equals unity, we are left with $M-1$
bin probabilities: $\pi_1, \pi_2, \ldots, \pi_{M-1}$, each
describing the probability that samples will be drawn from each of
the $M$ bins. The normalization condition requires that $\pi_M =
1-\sum_{k=1}^{M-1}{\pi_k}$. For simplicity, we assume that the bin
alignment is fixed so that extreme data points define the edges of the extreme bins.

\subsection{The Likelihood of the Piecewise-Constant Model}
The likelihood function is a probability density that when
multiplied by $dx$ describes the probability that a datum
$d_n$ is found to have a value in the infinitesimal range between some number
$x$ and $x + dx$. Since we have assumed that there is no additional measurement uncertainty associated with each datum, the likelihood that $d_n$ will have a value between $x$ and $x + dx$ falling within
the $k^{th}$ bin is given the uniform probability density
in the region defined by that bin
\begin{equation}
p(d_n | \pi_k, M, I) = h_k = \frac{\pi_k}{v_k}
\end{equation}
where $I$ represents our prior knowledge about the problem, which
includes the range of the data and the bin alignment.  For equal
width bins, the likelihood density reduces to
\begin{equation}
p(d_n | \pi_k, M, I) = \frac{M}{V}\pi_k.
\end{equation}
For $N$ independently sampled data points, the joint likelihood is
given by
\begin{equation} \label{eq:likelihood}
p(\underline{d} | \underline{\pi}, M, I) =
\biggl(\frac{M}{V}\biggr)^N \pi_1^{n_1} \pi_2^{n_2} \ldots
\pi_{M-1}^{n_{M-1}} \pi_{M}^{n_M}
\end{equation}
where $\underline{d} = \{d_1, d_2, \ldots, d_N\}$,
$\underline{\pi} = \{\pi_1, \pi_2, \ldots, \pi_{M-1}\}$, and the
$n_i$ are the number of data points in the $i^{th}$ bin. Equation
(\ref{eq:likelihood}) is data-dependent and describes the
likelihood that the hypothesized piecewise-constant model accounts
for the data. Individuals who recognize this as having the form of
the multinomial distribution may be tempted to include its
familiar normalization factor. However, it is important to note
that this likelihood function is properly normalized as is, which
we now demonstrate.  For a single datum $d$, the likelihood
that it will take the value $x$ is
\begin{equation}
p(d=x | \underline{\pi}, M, I) = \frac{1}{v} \sum_{k =
1}^{M}{\pi_k~\Pi(x_{k-1}, x, x_k)},
\end{equation}
where we have written $v = \frac{V}{M}$. Multiplying the
probability density by $dx$ to get the probability and integrating
over all possible values of $x$ we have
\begin{eqnarray}
\int_{-\infty}^{\infty}{dx~p(d=x | \underline{\pi}, M, I)}
&& = \int_{-\infty}^{\infty}{dx~\frac{1}{v} \sum_{k = 1}^{M}{\pi_k~\Pi(x_{k-1}, x, x_k)}}\nonumber\\
&& = \frac{1}{v} \sum_{k = 1}^{M}{\int_{-\infty}^{\infty}{dx~\pi_k~\Pi(x_{k-1}, x, x_k)}}\nonumber\\
&& = \frac{1}{v} \sum_{k = 1}^{M}{\pi_k~v}\nonumber\\
&& = \sum_{k = 1}^{M}{\pi_k}\nonumber\\
&& = 1.
\end{eqnarray}

\subsection{The Prior Probabilities}
For the prior probability of the number of bins, we assign a
uniform density
\begin{equation} \label{eq:prior-for-M}
p(M | I) =
   \left\{ \begin{array}{rl}
       C^{-1} & \mbox{if}~~1 \leq M \leq C\\
       0 & \mbox{otherwise} \end{array} \right.
\end{equation}
where $C$ is the maximum number of bins to be considered. This
could reasonably be set to the range of the data divided by
smallest non-zero distance between any two data points.

We assign a non-informative prior for the bin parameters $\pi_1,
\pi_2, \ldots, \pi_{M-1}$, the possible values of which lie within
a simplex defined by the corners of an $M$-dimensional hypercube
with unit side lengths
\begin{equation}
p(\underline{\pi} | M, I) =
\frac{\Gamma\bigl(\frac{M}{2}\bigr)}{\Gamma\bigl(\frac{1}{2}\bigr)^M}
\biggl[\pi_1 \pi_2 \cdots \pi_{M-1}
\biggl(1-\sum_{i=1}^{M-1}{\pi_i}\biggr)\biggr]^{-1/2}.
\label{eq:prior-for-pi's}
\end{equation}
Equation (\ref{eq:prior-for-pi's}) is the Jeffreys's prior for the
multinomial likelihood (\ref{eq:likelihood})
\cite{Jeffreys:1961,Box&Tiao:1992,Berger&Bernardo:1992}, and has
the advantage in that it is also the conjugate prior to the
multinomial likelihood.  The result is that the posterior probability is a Dirichlet-multinomial distribution, which is widely used in machine learning \cite{Bishop:2006}.  A similar posterior probability is used by Endres and Foldiak \cite{Endres&Foldiak:2005} to solve the more general problem of variable-width bin models.

\subsection{The Posterior Probability}
Using Bayes' Theorem, the posterior probability of the histogram
model is proportional to the product of the priors and the
likelihood
\begin{equation}
p(\underline{\pi}, M | \underline{d}, I) \propto p(\underline{\pi}
| M, I)~p(M | I)~p(\underline{d} | \underline{\pi}, M, I).
\end{equation}
Substituting (\ref{eq:likelihood}), (\ref{eq:prior-for-M}), and
(\ref{eq:prior-for-pi's}) gives the joint posterior probability
for the piecewise-constant density model
\begin{eqnarray}
&& p(\underline{\pi}, M | \underline{d}, I) \propto
\biggl(\frac{M}{V}\biggr)^N
\frac{\Gamma\bigl(\frac{M}{2}\bigr)}{\Gamma\bigl(\frac{1}{2}\bigr)^M}
\times \\ \label{eq:joint-posterior} && ~~~~~ \times
\pi_1^{n_1-\frac{1}{2}} \pi_2^{n_2-\frac{1}{2}} \ldots
\pi_{M-1}^{n_{M-1}-\frac{1}{2}}
\biggl(1-\sum_{i=1}^{M-1}{\pi_i}\biggr)^{n_M-\frac{1}{2}},\nonumber
\end{eqnarray}
where $p(M|I)$ is absorbed into the implicit proportionality
constant with the understanding that we will only consider a
reasonable range of bin numbers.

The goal is to obtain the posterior probability for the number of
bins $M$. To do this we integrate the joint posterior over all
possible values of $\pi_1, \pi_2, \ldots, \pi_{M-1}$ in the
simplex. While the result (\ref{eq:posterior-for-M}) is well-known \cite{Hutter:2001, Bishop:2006}, it is instructive to see how such integrations can be handled.  The expression we desire is written as a series of nested
integrals over the $M-1$ dimensional parameter space of bin
probabilities
\begin{align} \label{eq:nonsimplified-nested-integrals}
p(M | \underline{d}, I) \propto & \biggl(\frac{M}{V}\biggr)^N
\frac{\Gamma\bigl(\frac{M}{2}\bigr)}{\Gamma\bigl(\frac{1}{2}\bigr)^M}~
\int^{1}_{0} {d\pi_1~\pi_1^{n_1-\frac{1}{2}}} \int^{1-\pi_1}_{0}
{d\pi_2~\pi_2^{n_2-\frac{1}{2}}} \ldots\\
& \ldots \int^{(1-\sum_{i=1}^{M-2}{\pi_i})}_{0}
{d\pi_{M-1}~\pi_{M-1}^{n_{M-1}-\frac{1}{2}}
\biggl(1-\sum_{i=1}^{M-1}{\pi_i}\biggr)^{n_M-\frac{1}{2}}}.\nonumber
\end{align}
In order to write this more compactly, we first define
\begin{eqnarray}
 a_1 & = & 1\\
 a_2 & = & 1-\pi_1 \notag \\
 a_3 & = & 1-\pi_1-\pi_2 \notag \\
 \vdots \notag \\
 a_{M-1} & = & 1-\sum_{k=1}^{M-2}{\pi_k} \notag
\end{eqnarray}
and note the recursion relation
\begin{equation}
a_k = a_{k-1} - \pi_{k-1}.\label{eqn:recursion}
\end{equation}
These definitions greatly simplify the sum in the last term as
well as the limits of integration
\begin{align} \label{eq:nestedints}
p(M | \underline{d}, I) \propto & \biggl(\frac{M}{V}\biggr)^N
\frac{\Gamma\bigl(\frac{M}{2}\bigr)}{\Gamma\bigl(\frac{1}{2}\bigr)^M}~
\int^{a_1}_{0} {d\pi_1~\pi_1^{n_1-\frac{1}{2}}} \int^{a_2}_{0}
{d\pi_2~\pi_2^{n_2-\frac{1}{2}}} \ldots\\
& \ldots \int^{a_{M-1}}_{0}
{d\pi_{M-1}~\pi_{M-1}^{n_{M-1}-\frac{1}{2}}
(a_{M-1}-\pi_{M-1})^{n_M-\frac{1}{2}}}.\nonumber
\end{align}
To solve the set of nested integrals in
(\ref{eq:nonsimplified-nested-integrals}), consider the general
integral
\begin{equation}
I_k = \int^{a_k}_{0} {d\pi_k~\pi_k^{n_k-\frac{1}{2}}
(a_k-\pi_k)^{b_k}}.
\end{equation}
This integral can be re-written as
\begin{equation}
I_k = a_k^{b_k} \int^{a_k}_{0} {d\pi_k~\pi_k^{n_k-\frac{1}{2}}
\biggl( 1 - \frac{\pi_k}{a_k} \biggr)^{b_k}}.
\end{equation}
Setting $\displaystyle u = \frac{\pi_k}{a_k}$ we have
\begin{eqnarray}
I_k & = & a_k^{b_k} \int^{1}_{0} {du~a_k^{n_k+\frac{1}{2}}
u^{n_k-\frac{1}{2}} (1 - u)^{b_k}} \nonumber\\
& = & a_k^{b_k+n_k+\frac{1}{2}} \int^{1}_{0}
{du~u^{n_k-\frac{1}{2}}
(1 - u)^{b_k}}, \nonumber \\
& = & a_k^{b_k+n_k+\frac{1}{2}} B(n_k + \frac{1}{2}, b_k +
1)\label{eq:intsoln}
\end{eqnarray}
where $B(\cdot)$ is the Beta function with
\begin{equation} \label{eq:beta-ito-gamma}
B(n_k + \frac{1}{2}, b_k + 1) = \frac{\Gamma\biggl( n_k +
\frac{1}{2} \biggr) \Gamma(b_k + 1)}{\Gamma\biggl( n_k +
\frac{1}{2} + b_k + 1 \biggr)}.
\end{equation}
To solve all of the integrals we rewrite $a_k$ in
(\ref{eq:intsoln}) using the recursion formula
(\ref{eqn:recursion})
\begin{equation}
I_k = (a_{k-1}-\pi_{k-1})^{b_k+n_k+\frac{1}{2}} B(n_k +
\frac{1}{2}, b_k + 1).
\end{equation}
By defining
\begin{eqnarray} \label{eq:b-recursion}
 b_{M-1} & = & n_M-\frac{1}{2}\\
 b_{k-1} & = & b_k+n_k+\frac{1}{2}\notag
\end{eqnarray}
we find
\begin{eqnarray}
 b_{1} & = & N-n_1+\frac{M}{2}-\frac{3}{2}.
\end{eqnarray}
Finally, integrating (\ref{eq:nestedints}) gives
\begin{equation} \label{eq:posterior-for-M-with-Betas}
p(M | \underline{d}, I) \propto \biggl(\frac{M}{V}\biggr)^N
\frac{\Gamma\bigl(\frac{M}{2}\bigr)}{\Gamma\bigl(\frac{1}{2}\bigr)^M}~
\prod_{k=1}^{M-1}{B(n_k + \frac{1}{2}, b_k + 1)},\nonumber
\end{equation}
which can be simplified further by expanding the Beta functions
using (\ref{eq:beta-ito-gamma})
\begin{eqnarray}
p(M | \underline{d}, I) \propto \biggl(\frac{M}{V}\biggr)^N &&
\frac{\Gamma\bigl(\frac{M}{2}\bigr)}{\Gamma\bigl(\frac{1}{2}\bigr)^M} \times \\
&&\frac{\Gamma(n_1 + \frac{1}{2}) \Gamma(b_1 + 1)}{\Gamma(n_1 +
\frac{1}{2} + b_1 + 1)} \times \frac{\Gamma(n_2 + \frac{1}{2})
\Gamma(b_2 + 1)}{\Gamma(n_2 + \frac{1}{2} + b_2 + 1)} \times \nonumber\\
&&\ldots \times \frac{\Gamma(n_{M-1} + \frac{1}{2}) \Gamma(b_{M-1}
+ 1)}{\Gamma(n_{M-1} + \frac{1}{2} + b_{M-1} + 1)} \nonumber
\end{eqnarray}
Using the recursion relation (\ref{eq:b-recursion}) for the $b_k$,
we see that the general term $\Gamma(b_k+1)$ in each numerator,
except the last, cancels with the denominator in the following
term.  This leaves
\begin{equation}
p(M | \underline{d}, I) \propto \biggl(\frac{M}{V}\biggr)^N
\frac{\Gamma\bigl(\frac{M}{2}\bigr)}{\Gamma\bigl(\frac{1}{2}\bigr)^M}~
\frac{\prod_{k=1}^{M}{\Gamma(n_k+\frac{1}{2})}}{\Gamma(n_1+b_1+\frac{3}{2})},
\end{equation}
where we have used (\ref{eq:b-recursion}) to observe that
$\Gamma(b_{M-1}+1) = \Gamma(n_M+1/2)$. Last, again using the
recursion relation in (\ref{eq:b-recursion}) we find that $b_1 =
N-n_1+\frac{M}{2}-\frac{3}{2}$, which results in our marginal
posterior probability
\begin{equation} \label{eq:posterior-for-M}
p(M | \underline{d}, I) \propto \biggl(\frac{M}{V}\biggr)^N
\frac{\Gamma\bigl(\frac{M}{2}\bigr)}{\Gamma\bigl(\frac{1}{2}\bigr)^M}~
\frac{\prod_{k=1}^{M}{\Gamma(n_k+\frac{1}{2})}}{\Gamma(N+\frac{M}{2})}.
\end{equation}
The normalization of this posterior probability density depends on
the actual data used.  For this reason, we will work with the
un-normalized posterior, and shall refer to its values as relative
posterior probabilities.

In optimization problems, it is often easier to maximize the
logarithm of the posterior
\begin{align} \label{eq:log-posterior-for-M}
\log{p(M | \underline{d}, I)} = N \log{M} +
\log{\Gamma\biggl(\frac{M}{2}\biggr)} + \\ \nonumber ~~~~~ - M
\log{\Gamma\biggl(\frac{1}{2}\biggr)}
-\log{\Gamma\biggl(N+\frac{M}{2}\biggr)} + \\ \nonumber ~~~~~ +
\sum_{k=1}^{M}{\log{\Gamma\biggl(n_k+\frac{1}{2}\biggr)}} + K,
\end{align}
where $K$ represents the sum of the volume term and the logarithm
of the implicit proportionality constant. The optimal number of
bins $\hat{M}$ is found by identifying the mode of the logarithm
of the marginal posterior
\begin{equation}
\hat{M} = \underset{M}{\textrm{arg max}} \{\log{p(M |
\underline{d}, I)}\}.
\end{equation}
Such a result is reassuring, since it is independent of the order
in which the bins are counted.  Many software packages are
equipped to quickly compute the log of the gamma function.
However, for more basic implementations, the following definitions
from Abramowitz and Stegun \cite{Abramowitz&Stegun:1972}
can be used for integer~$m$ .
\begin{align}
& \log{\Gamma(m)} = \sum_{k=1}^{m-1}{\log{k}}\\
& \log{\Gamma \biggl( m+\frac{1}{2} \biggr)} =
\frac{1}{2}\log{\pi} - n\log{2} + \sum_{k=1}^{m}{\log{(2k-1)}}
\end{align}
Equation (\ref{eq:log-posterior-for-M}) allows one to easily
identify the number of bins $M$ which optimize the posterior. We
call this technique the \texttt{optBINS} algorithm and provide the
\texttt{Matlab} code in the Appendix.

\section{The Posterior Probability for the Bin Height}
In order to obtain the posterior probability for the probability
mass of a particular bin, we begin with the joint posterior
(\ref{eq:joint-posterior}) and integrate over all the other bin
probability masses.  Since we can consider the bins in any order,
the resulting expression is similar to the multiple nested
integral in (\ref{eq:nonsimplified-nested-integrals}) except that
the integral for one of the $M-1$ bins is not performed.  Treating
the number of bins as a given, we can use the product rule to get
\begin{equation}
p(\underline{\pi} | \underline{d}, M, I) =
\frac{p(\underline{\pi}, M | \underline{d}, I)}{p(M |
\underline{d}, I)}
\end{equation}
where the numerator is given by (\ref{eq:joint-posterior}) and the
denominator by (\ref{eq:posterior-for-M}). Since the bins can be
treated in any order, we derive the marginal posterior for the
first bin and generalize the result for the $k^{th}$ bin.  The
marginal posterior is
\begin{align}
& p(\pi_1 | \underline{d}, M, I) = \frac{(\frac{M}{V})^N
\frac{\Gamma\bigl(\frac{M}{2}\bigr)}{\Gamma\bigl(\frac{1}{2}\bigr)^M}}{p(M
| \underline{d},I)} \pi_1^{n_1-\frac{1}{2}}~~~~\times \\ \nonumber
&~~~~~~ \times \int^{a_2}_{0} {d\pi_2~\pi_2^{n_2-\frac{1}{2}}}
\int^{a_3}_{0} {d\pi_3~\pi_3^{n_3-\frac{1}{2}}} \ldots\\ \nonumber
&~~~~~\ldots \int^{a_{M-1}}_{0}
{d\pi_{M-1}~\pi_{M-1}^{n_{M-1}-\frac{1}{2}}
(a_{M-1}-\pi_{M-1})^{n_M-\frac{1}{2}}}.
\end{align}
Evaluating the integrals and substituting
(\ref{eq:posterior-for-M-with-Betas}) into the denominator we get
\begin{equation}
p(\pi_1 | \underline{d}, M, I) =
\frac{\prod_{k=2}^{M-1}{B(n_k+\frac{1}{2},
b_k+1)}}{\prod_{k=1}^{M-1}{B(n_k+\frac{1}{2}, b_k+1)}}
\pi_1^{n_1-\frac{1}{2}} (1-\pi_1)^{b_1}.
\end{equation}
Cancelling terms and explicitly writing $b_1$, the marginal
posterior for $\pi_1$ is
\begin{align}
& p(\pi_1 | \underline{d}, M, I) = \\
&~~\frac{\Gamma(N+\frac{M}{2})}{\Gamma(n_1+\frac{1}{2})
\Gamma(N-n_1+\frac{M-1}{2})} \pi_1^{n_1-\frac{1}{2}}
(1-\pi_1)^{N-n_1+\frac{M-3}{2}},  \nonumber
\end{align}
which can easily be verified to be normalized by integrating
$\pi_1$ over its entire possible range from 0 to 1.  Since the
bins can be considered in any order, this is a general result for
the $k^{th}$ bin
\begin{align} \label{eq:posterior-bin-height}
& p(\pi_k | \underline{d}, M, I) = \\
&~~\frac{\Gamma(N+\frac{M}{2})}{\Gamma(n_k+\frac{1}{2})
\Gamma(N-n_k+\frac{M-1}{2})} \pi_k^{n_k-\frac{1}{2}}
(1-\pi_k)^{N-n_k+\frac{M-3}{2}}.  \nonumber
\end{align}

The mean bin probability mass can be found from its expectation
\begin{equation}
\langle\pi_k\rangle~~=~~\int_0^1{d\pi_k~\pi_k~p(\pi_k |
\underline{d}, M, I)},
\end{equation}
which substituting (\ref{eq:posterior-bin-height}) gives
\begin{align}
\langle\pi_k\rangle~~=~~&
\frac{\Gamma(N+\frac{M}{2})}{\Gamma(n_k+\frac{1}{2})
\Gamma(N-n_k+\frac{M-1}{2})}~~\times \\
&~~~~~~~~~~~~~~~\int_0^1{d\pi_k~\pi_k^{n_k+\frac{1}{2}}
(1-\pi_k)^{N-n_k+\frac{M-3}{2}}}. \nonumber
\end{align}
The integral again gives a Beta function, which when written in
terms of Gamma functions is
\begin{align}
\langle\pi_k\rangle~~=~~&
\frac{\Gamma(N+\frac{M}{2})}{\Gamma(n_k+\frac{1}{2})
\Gamma(N-n_k+\frac{M-1}{2})} \times \\
&~~~~~~~~~~~~~~~\frac{\Gamma(n_k+\frac{3}{2})
\Gamma(N-n_k+\frac{M-1}{2})}{\Gamma(N+\frac{M}{2}+1)}. \nonumber
\end{align}
Using the fact that $\Gamma(x+1) = x\Gamma(x)$ and cancelling like
terms, we find that
\begin{equation} \label{eq:mean-of-bin-heights}
\langle\pi_k\rangle = \frac{n_k+\frac{1}{2}}{N+\frac{M}{2}}.
\end{equation}
The mean probability density for bin $k$ (the bin height) is
simply
\begin{equation} \label{eq:mean-of-prob-density}
\mu_k = \langle h_k \rangle = \frac{\langle \pi_k \rangle}{v_k} =
\biggl(\frac{M}{V}\biggr)
\biggl(\frac{n_k+\frac{1}{2}}{N+\frac{M}{2}}\biggr).
\end{equation}
It is an interesting result that bins with no counts still have a
non-zero probability. This makes sense since no lack of evidence
can ever prove conclusively that an event occurring in a given bin
is impossible---just less probable. The Jeffrey's prior
effectively places one-half of a datum in each bin.

The variance of the probability mass of the $k^{th}$ bin is found similarly by
\begin{equation}
\sigma_k^2 = \biggl(\frac{M}{V}\biggr)^2
\bigl(\langle\pi_k^2\rangle - \langle\pi_k\rangle^2\bigr),
\end{equation}
which gives
\begin{equation} \label{eq:variance-of-prob-density}
\sigma_k^2 = \biggl(\frac{M}{V}\biggr)^2
\biggl(\frac{(n_k+\frac{1}{2})(N-n_k+\frac{M-1}{2})}
{(N+\frac{M}{2}+1)(N+\frac{M}{2})^2}\biggr).
\end{equation}

Thus, given the optimal number of bins found by maximizing
(\ref{eq:log-posterior-for-M}), the mean and variance of the bin
probabilities are found from (\ref{eq:mean-of-prob-density}) and
(\ref{eq:variance-of-prob-density}), which allow us to construct
an explicit histogram model of the probability density and perform
computations replete with proper error analysis. Note that in the case where there
is one bin (\ref{eq:variance-of-prob-density}) gives a zero
variance.

\begin{figure}
\centering
\makebox{\includegraphics[scale=0.80]{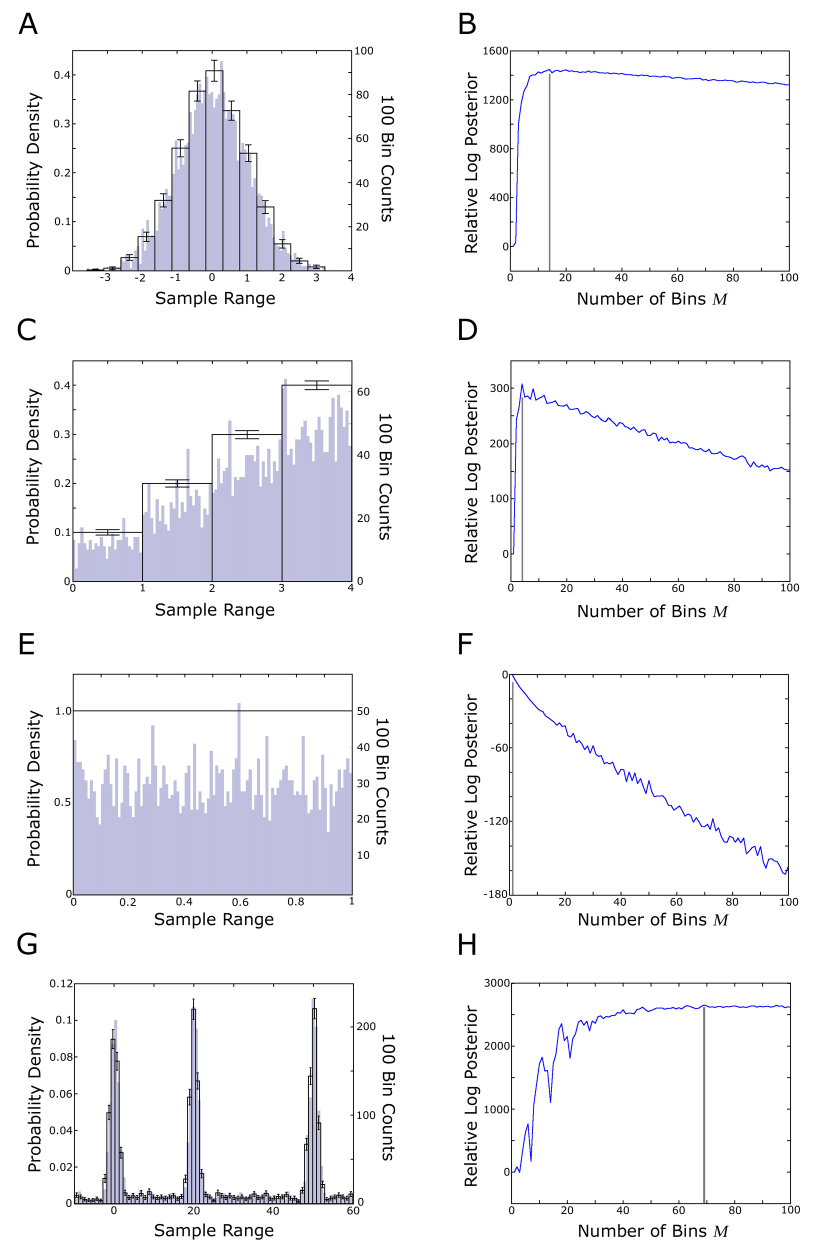}}
\caption{To demonstrate the technique, 1000 samples were sampled
from four different probability density functions. (A)
The optimal piecewise-constant model for 1000 samples drawn from a Gaussian
density function is superimposed over a 100-bin histogram that
shows the distribution of data samples. (B) The relative log posterior
probability of the number of bins peaks at 14 bins for these 1000
data sampled from the Gaussian density. (C) Samples shown are from
a 4-step piecewise-constant density function. The relative log posterior peaks at four bins (D) indicating that the method correctly detects the four-step structure.  (E) These data were sampled from a uniform
density as verified by the relative log posterior probability shown in (F), which starts at a maximum value of one and decreases with increasing numbers of bins. (G) Here we demonstrate a more complex example---three Gaussian peaks plus a uniform
background. (H) The posterior, which peaks at 52 bins,
demonstrates clearly that the data themselves support this
detailed picture of the pdf.} \label{fig:01}
\end{figure}

\section{Results}
\subsection{Demonstration using One-Dimensional Histograms} \label{sec:demo}
In this section we demonstrate the utility of this method for
determining the optimal number of bins in a piecewise-constant density model by applying this method to
several different data sets.  Note that since it is computationally costly to marginalize the posterior probability (\ref{eq:posterior-for-M}) to obtain the appropriate normalization factor, the analyses below rely on the un-normalized posterior probability, the logarithms of which will be referred to the \textit{relative log posterior}.  We consider four different test cases where we have sampled 1000 data points from each of the four different probability density functions.

The first example considers a Gaussian probability density $\mathcal{N}(0,1)$. The optimal piecewise-constant density model for the 1000 data points sampled from this distribution is shown in Figure 1A, where it is superimposed over a 100-bin histogram that better illustrates the locations of the sampled points. Figure 1B shows that the relative log posterior probability (\ref{eq:log-posterior-for-M}) peaks at 14 bins.  Note that the bin heights for the piecewise-constant density model are determined from (\ref{eq:mean-of-bin-heights}); whereas the bin heights of the 100-bin histogram illustrating the data samples are proportional to the counts.  For this reason, the two pictures are not directly comparable.

The second example considers a 4-step constant piecewise density.  Figure 1C shows the optimal binning for the 1000 sampled data points.  The relative log posterior (Figure 1D) peaks at 4 bins, which indicates that the method correctly detects the 4-step structure.

A uniform density is used to sample 1000 data points in the third example.  Figures 1E and 1F, demonstrate that samples drawn from a uniform density were best described by a single bin. This result is significant, since entropy estimates computed from these data would be biased if multiple bins were used to describe the distribution of the sampled data.

Last, we consider a density function that consists of a mixture of three sharply-peaked Gaussians with a uniform background (Figure 1G). The posterior peaks at 52 bins indicating that the data warrant a detailed model (Figure 1H). The
spikes in the relative log posterior are due to the fact that the bin edges
are fixed. The relative log posterior is large at values of $M$ where the
bins happen to line up with the Gaussians, and small when they are
misaligned. This last example demonstrates one of the weaknesses
of the equal bin-width model, as many bins are needed to describe
the uniform density between the three narrow peaks.  In addition, the lack of an obvious peak indicates that there is a range of bin numbers that will result in reasonable models.

\begin{figure}
\centering
\makebox{\includegraphics[scale=0.85]{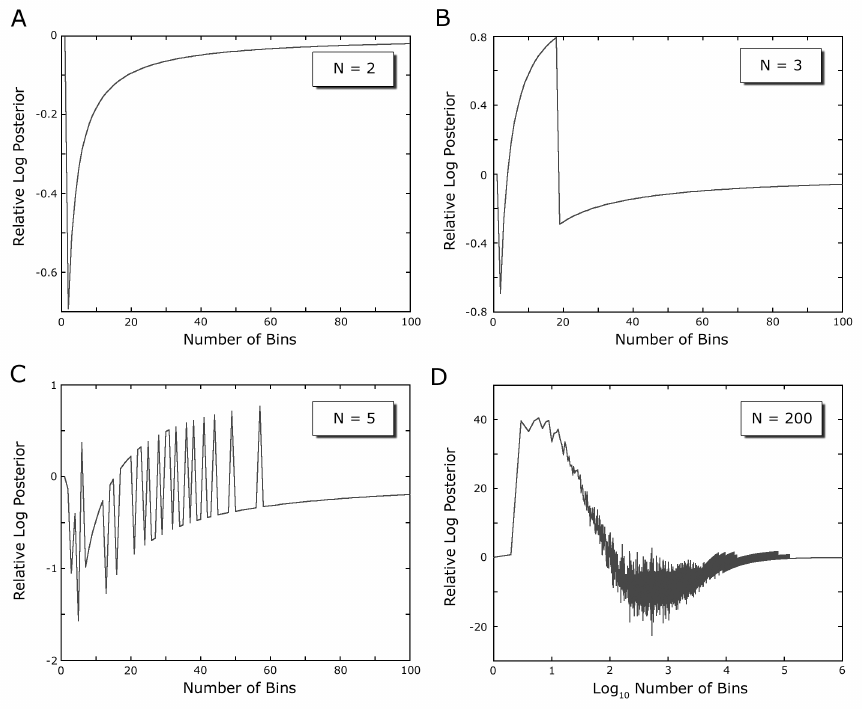}}
\caption{These figures demonstrate the behavior of the relative
log posterior for small numbers of samples. (A) With only $N=2$
samples, the log posterior is maximum when the samples are in the
same bin $M=1$. For $M>1$, the log posterior follows the function
described in (\ref{eq:two-samples}) in the text. (B) The relative
log posterior is slightly more complicated for $N=3$.  For $M=1$
all three points lie in the same bin.  As $M$ increases, two data
points are in one bin and the remaining datum point is in another
bin.  The functional form is described by (\ref{eq:N=3,2|1}).
Eventually, all three data points lie in separate bins and the
relative log posterior is given by (\ref{eq:N=3,1|1|1}). (C) The
situation is more complicated still for $N=5$ data points.  As $M$
increases, a point is reached when, depending on the particular
value of $M$, the points will be in separate bins.  As $M$ changes
value, two points may again fall into the same bin.  This gives
rise to this oscillation in the log posterior.  Once all points
are in separate bins, the behavior follows a well-defined
functional form (\ref{eq:M>N}). (D) This plot shows the behavior
for a large number of data points $N=200$. The log posterior now
displays a more well-defined mode indicating that there is a
well-defined optimal number of bins.  As $M$ approaches 10000 to
100000 bins, one can see some of the oscillatory behavior
demonstrated in the small $N$ cases.}\label{fig:small-samples}
\end{figure}

\section{Effects of Small Sample Size}
\subsection{Small Samples and Asymptotic Behavior}
It is instructive to observe how this algorithm behaves in
situations involving small sample sizes. We begin by considering
the extreme case of two data points $N=2$.  In the case of a
single bin, $M=1$, the posterior probability reduces to
\begin{eqnarray}
p(M=1 | d_1, d_2, I) & \propto & M^N
\frac{\Gamma\bigl(\frac{M}{2}\bigr)}{\Gamma\bigl(\frac{1}{2}\bigr)^M}~
\frac{\prod_{k=1}^{M}{\Gamma(n_k+\frac{1}{2})}}{\Gamma(N+\frac{M}{2})}
\nonumber \\
& \propto & 1^2
\frac{\Gamma\bigl(\frac{1}{2}\bigr)}{\Gamma\bigl(\frac{1}{2}\bigr)^1}
\frac{\Gamma\bigl(2+\frac{1}{2}\bigr)}{\Gamma\bigl(2+\frac{1}{2}\bigr)}
= 1,
\end{eqnarray}
so that the log posterior is zero. For $M>1$, the two data points
lie in separate bins, resulting in
\begin{eqnarray}
\label{eq:two-samples} p(M | d_1, d_2, I) & \propto & M^N
\frac{\Gamma\bigl(\frac{M}{2}\bigr)}{\Gamma\bigl(\frac{1}{2}\bigr)^M}~
\frac{\prod_{k=1}^{M}{\Gamma(n_k+\frac{1}{2})}}{\Gamma(N+\frac{M}{2})}
\nonumber \\
& \propto & M^2
\frac{\Gamma\bigl(\frac{M}{2}\bigr)}{\Gamma\bigl(\frac{1}{2}\bigr)^M}~
\frac{\Gamma(1+\frac{1}{2})^2
\Gamma(\frac{1}{2})^{M-2}}{\Gamma(2+\frac{M}{2})}\nonumber \\
& \propto & M^2
\frac{\Gamma(\frac{3}{2})^2}{\Gamma\bigl(\frac{1}{2}\bigr)^2}~
\frac{\Gamma\bigl(\frac{M}{2}\bigr)}{\Gamma(2+\frac{M}{2})}\nonumber \\
& \propto & \frac{1}{2} \cdot \frac{M}{1+\frac{M}{2}}.
\end{eqnarray}
Figure \ref{fig:small-samples}A shows the log posterior which
starts at zero for a single bin, drops to $\log(\frac{1}{2})$ for
$M=2$ and then increases monotonically approaching zero in the
limit as $M$ goes to infinity.  The result is that a single bin is
the most probable solution for two data points.

For three data points in a single bin ($N=3$ and $M=1$), the
posterior probability is one, resulting in a log posterior of
zero.  In the $M>1$ case where there are two data points in one
bin and one datum point in another, the posterior probability is
\begin{equation}\label{eq:N=3,2|1}
p(M | d_1, d_2, d_3, I) \propto \frac{3}{4} \cdot
\frac{M^2}{(2+\frac{M}{2})(1+\frac{M}{2})},
\end{equation}
and for each point in a separate bin we have
\begin{equation}\label{eq:N=3,1|1|1}
p(M | d_1, d_2, d_3, I) \propto \frac{1}{4} \cdot
\frac{M^2}{(2+\frac{M}{2})(1+\frac{M}{2})}.
\end{equation}
While the logarithm of the un-normalized posterior in (\ref{eq:N=3,2|1}) can be
greater than zero, as $M$ increases, the data points eventually
fall into separate bins. This causes the posterior to change from
(\ref{eq:N=3,2|1}) to (\ref{eq:N=3,1|1|1}) resulting in a dramatic
decrease in the logarithm of the posterior, which then
asymptotically increases to zero as $M \rightarrow \infty$. This
behavior is shown in Figure \ref{fig:small-samples}B.

More rich behavior can be seen in the case of $N=5$ data points.
The results again (Figure \ref{fig:small-samples}C)  depend on the
relative positions of the data points with respect to one another.
In this case the posterior probability switches between two types
of behavior as the number of bins increase depending on whether
the bin positions force two data points together in the same bin
or separate them into two bins.  The ultimate result is a
ridiculous \emph{maximum a posteriori} solution of 57 bins.
Clearly, for a small number of data points, the mode depends
sensitively on the relative positions of the samples in a way that
is not meaningful.  In these cases there are too few data points
to model a density function.

With a larger number of samples, the posterior probability shows a
well-defined mode indicating a well-determined optimal number of
bins.  In the general case of $M > N$ where each of the $N$ data
points is in a separate bin, we have
\begin{equation}\label{eq:M>N}
p(M | \underline{d}, I) \propto \biggl(\frac{M}{2}\biggr)^N
\frac{\Gamma\bigl(\frac{M}{2}\bigr)}{\Gamma\bigl(N+\frac{M}{2}\bigr)},
\end{equation}
which again results in a log posterior that asymptotically
approaches zero as $M \rightarrow \infty$.  Figure
\ref{fig:small-samples}D demonstrates these two effects for
$N=200$.  This also can be compared to the log posterior for
$1000$ Gaussian samples in Figure \ref{fig:01}B.

\begin{figure}
\centering
\makebox{\includegraphics[scale=0.35]{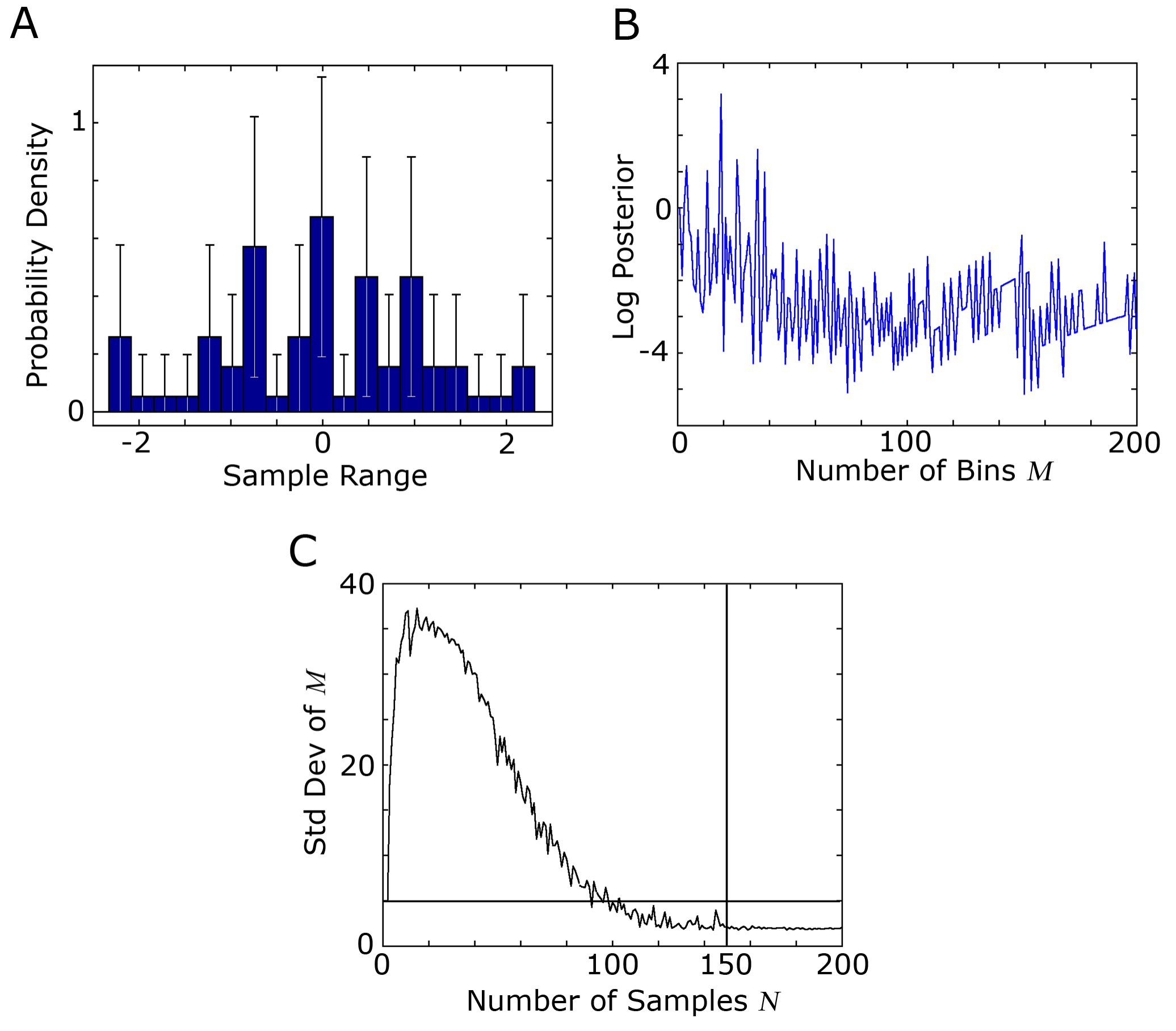}}
\caption{(A) An optimal density model ($M = 19$) for $N = 30$ data
points sampled from a Gaussian distribution. The fact that the
error bars on the bin probabilities are as large as the
probabilities themselves indicates that this is a poor estimate.
(B) The log posterior probability for the number of bins possesses
no well-defined peak, and is instead reminiscent of noise.  (C)
This plot shows the standard deviation of the estimated number of
bins $M$ for $1000$ data sets of $N$ points, ranging from $2$ to
$200$, sampled from a Gaussian distribution.  The standard
deviation stabilizes around $\sigma_M = 2$ bins for $N > 150$
indicating the inherent level of uncertainty in the problem.  This
suggests that one requires at least $150$ data points to
consistently perform such probability density estimates, and can
perhaps get by with as few as $100$ data points in some
cases.}\label{fig:sufficient}
\end{figure}

\subsection{Sufficient Data}
The investigation on the effects of small sample size in the previous section raises the
question as to how many data points are needed to estimate the
probability density function.  The general shape of a healthy log
posterior reflects a sharp initial rise to a well-defined peak,
and a gradual fall-off as the number of bins $M$ increases from
one (eg. Fig. \ref{fig:01}B, Fig. \ref{fig:small-samples}D).  With
small sample sizes, however, one finds that the bin heights have
large error bars (Figure \ref{fig:sufficient}A) so that $\mu_i
\simeq \sigma_i$, and that the log posterior is multi-modal
(Figure \ref{fig:sufficient}B) with no clear peak.

We tested our algorithm on data sets with $199$ different sample
sizes from $N = 2$ to $N = 200$.  One thousand data sets were
drawn from a Gaussian distribution for each value of $N$. The
standard deviation of the number of bins obtained for these 1000
data sets at a given value if $N$ was used as an indicator of the
stability of the solution.

Figure \ref{fig:sufficient}C shows a plot of the standard
deviation of the number of bins selected for the 1000 data sets at
each value of $N$. As we found above, with two data points, the
optimal solution is always one bin giving a standard deviation of
zero. This increases dramatically as the number of data points
increases, as we saw in our example with $N = 5$ and $M = 57$.
This peaks around $N = 15$ and slowly decreases as $N$ increases
further.  The standard deviation of the number of bins decreased
to $\sigma_M < 5$ for $N > 100$, and stabilized to $\sigma_M
\simeq 2$ for $N > 150$.

While 30 samples may be sufficient for estimating the mean and
variance of a density function known to be Gaussian, it is clear
that more samples are needed to reliably estimate the shape of an
unknown density function.  In the case where the data are
described by a Gaussian, it would appear that at least $150$ samples, are required to
accurately and consistently infer the shape of a one-dimensional density
function. By examining the shape of the log posterior, one can
easily determine whether one has sufficient data to estimate the
density function. In the event that there are too few samples to
perform such estimates, one can either incorporate additional
prior information or collect more data.

\section{Digitized Data} \label{sec:dig}
Due to the way that computers represent data, all data are
essentially represented by integers \cite{Bayman&Broadhurst}. In
some cases, the data samples have been intentionally rounded or
truncated, often to save storage space or transmission time.  It
is well-known that any non-invertible transformation, such as
rounding, destroys information.  Here we investigate how severe
losses of information due to rounding or truncation affects the
\texttt{optBINS} algorithm.

\begin{figure}
\centering
\makebox{\includegraphics[scale=0.30]{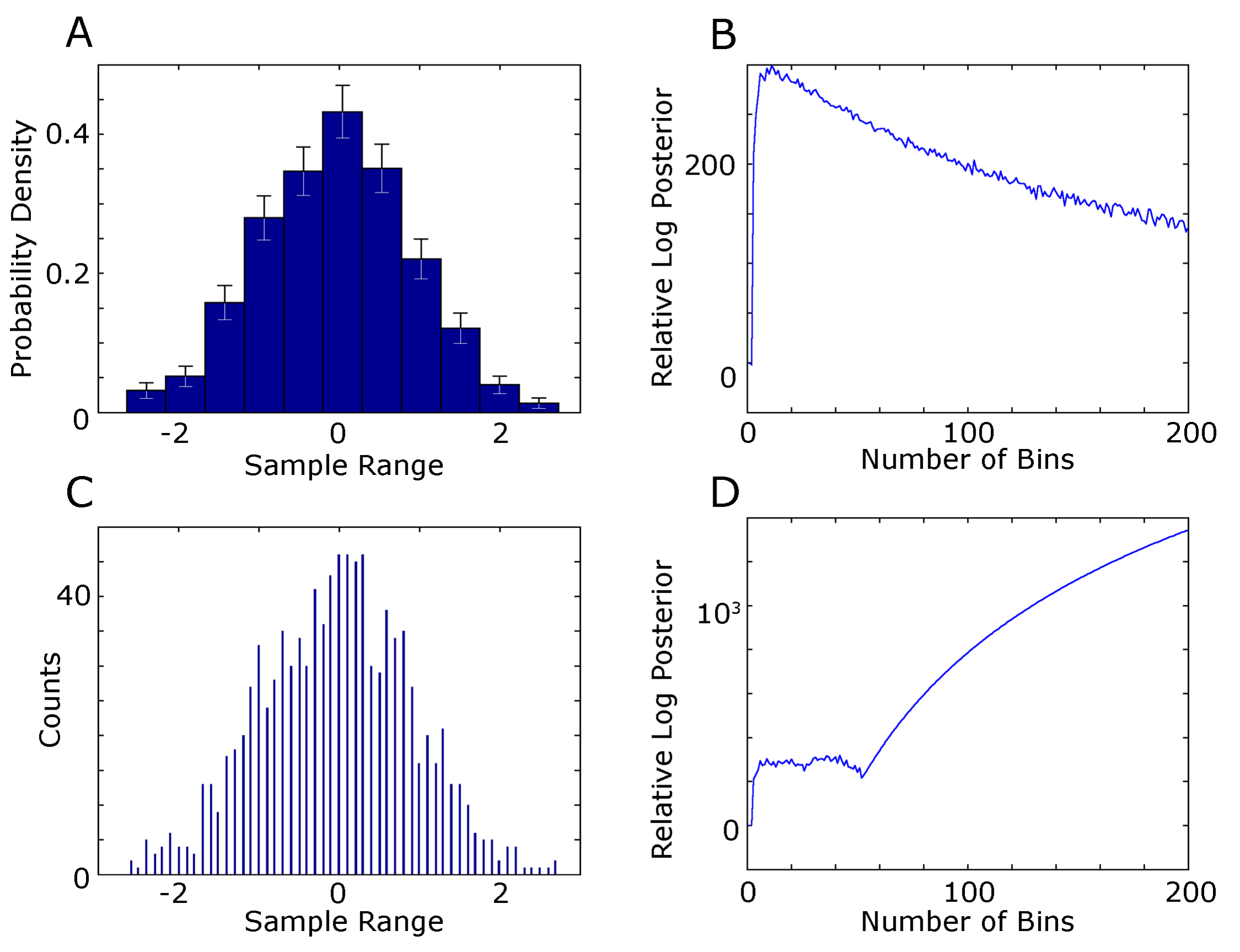}}
\caption{$N=1000$ data points were sampled from a Gaussian
distribution $\mathcal{N}(0,1)$. The top plots show (A) the
estimated density function using optimal binning and (B) the
relative log posterior, which exhibits a well-defined peak at
$M=11$ bins.  The bottom plots reflect the results using the same
data set after it has been rounded with $\Delta x = 0.1$ to keep
only the first decimal place.  (C) There is no optimal binning as
the algorithm identifies the discrete structure as being a more
salient feature than the overall Gaussian shape of the density
function.  (D) The relative log posterior displays no well-defined
peak, and in addition, for large numbers of $M$ displays a
monotonically increase given by (\ref{eq:M>N}) that asymptotes to
a positive value. This indicates that the data have been severely
rounded.}\label{fig:rounded}
\end{figure}

When data are digitized via truncation or rounding, the
digitization is performed so as to maintain a resolution that we
will denote by $\Delta x$.  That is, if the data set has values
that range from 0 to 1, and we represent these numbers with eight
bits, the minimum resolution we can maintain is $\Delta x = 1/2^8
= 1/256$.  For a sufficiently large data set (in this example
$N>256$) the pigeonhole principle indicates that it will be impossible to have a situation where each datum is in its
own bin when the number of bins is greater than a critical number,
$M > M_{\Delta x}$, where
\begin{equation}
M_{\Delta x} = \frac{V}{\Delta x},
\end{equation}
and $V$ is the range of the data considered (see Figure 5).  Once
$M > M_{\Delta x}$ the number of populated bins $P$ will remain
unchanged since the bin width $w$ for $M > M_{\Delta x}$ will be
smaller then the digitization resolution, $w < \Delta x$.

For all bin numbers $M > M_{\Delta x}$, there will be $P$
populated bins with populations $n_1, n_2, \ldots, n_P$.  This
leads to a form for the marginal posterior probability for $M$
(\ref{eq:posterior-for-M}) that depends only on the number of
instances of each discrete value that was recorded, $n_1, n_2,
\ldots, n_P$. Since these values do not vary for $M > M_{\Delta
x}$, the marginal posterior can be expressed solely as a function of
$M$
\begin{equation}
p(M | \underline{d}, I) \propto \biggl(\frac{M}{2}\biggr)^N
\frac{\Gamma\bigl(\frac{M}{2}\bigr)}{\Gamma\bigl(N+\frac{M}{2}\bigr)}
\cdot 2^N
\frac{\prod_{p=1}^{P}{\Gamma(n_p+\frac{1}{2})}}{\Gamma\bigl(\frac{1}{2}\bigr)^P},
\end{equation}
where the product over $p$ is over populated bins only. Comparing
this to (\ref{eq:M>N}), the function on the right-hand side
asymptotically approaches a value greater than one so that its
logarithm increases asymptotically to a value greater than zero.

As the number of bins $M$ increases, the point is reached where
the data can not be further separated; call this point $M_{crit}$.
In this situation, there are $n_p$ data points in the $p^{th}$ bin
and the posterior probability can be written as
\begin{equation}\label{eq:picket-fence-posterior}
p(M | \underline{d}, I) \propto \biggl(\frac{M}{2}\biggr)^N
\frac{\Gamma\bigl(\frac{M}{2}\bigr)}{\Gamma\bigl(N+\frac{M}{2}\bigr)}
\cdot \prod_{p = 1}^P {(2n_p-1)!!},
\end{equation}
where $!!$ denotes the double factorial \cite{Arfken:1985}. For $M > M_{crit}$, as $M
\rightarrow \infty$, the log posterior asymptotes to $\sum_{p =
1}^P {\log((2n_p-1)!!)}$, which can be further simplified to
\begin{equation}\label{eq:picket-fence-log-asymptote}
\sum_{p = 1}^P{\log((2n_p-1)!!)} = (P-N)\log(2) + \sum_{p =
1}^P{\sum_{s=n_p}^{2n_p-1}{\log s}}.
\end{equation}

This means that excessive truncation or rounding can be detected by comparing the the mode of $\log
p(M | \underline{d}, I)$ for $M < M_{crit}$ to
(\ref{eq:picket-fence-log-asymptote}) above.  If the latter is
larger, this indicates that the discrete nature of the data is a more significant
feature than the general shape of the underlying probability
density function.  When this is the case, a reasonable histogram
model of the density function can still be obtained by adding a
uniformly-distributed random number, with a range defined by the
resolution $\Delta x$, to each datum point
\cite{Bayman&Broadhurst}. While this will produce the best
histogram possible given the data, this will not recover the lost
information.

\begin{figure}
\centering
\makebox{\includegraphics[scale=0.8]{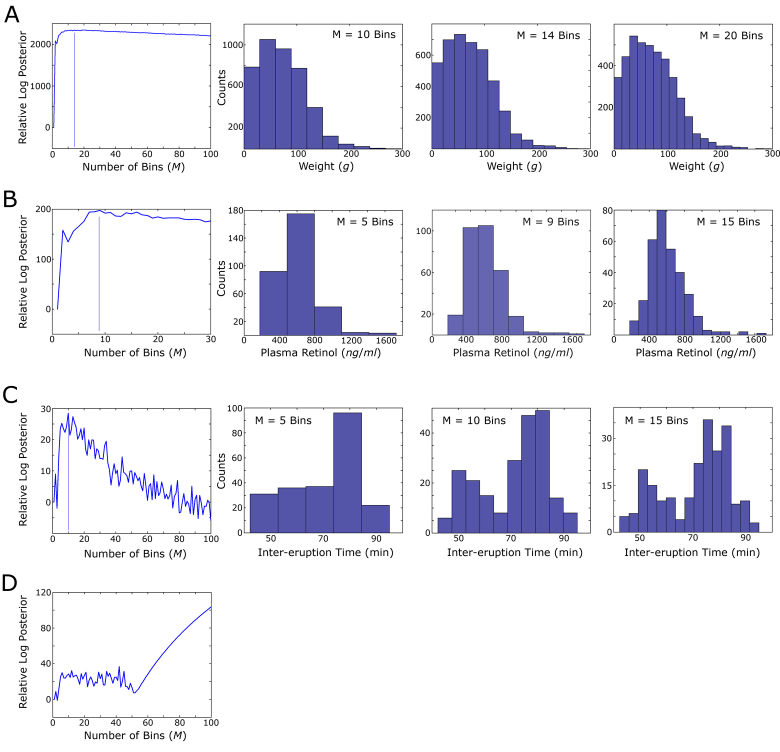}}
\caption{The \texttt{optBINS} algorithm is used to select the
number of bins for three real data sets.  The relative log
posterior probability of each of the three data sets is displayed
on the left where the optimal number of bins is the point at which
the global maximum occurs (see drop lines).  The center density
function in each row represents the optimally-binned model.  The
leftmost density function has too few bins and is not sufficiently
well-resolved, whereas the rightmost density function has too many
bins and is beginning to highlight irrelevant features. (A)
Abalone weights in grams from 4177 individuals \cite{Nash:1994}.
(B) Blood plasma retinol concentration in ng/ml measured from 315
individuals \cite{Nierenberg:1989}. (C) The Old Faithful data set
consisting of measurements of inter-eruption intervals rounded to
the nearest minute \cite{Azzalini&Bowman:1990}. In this data set,
we added a uniformly distributed random number from -0.5 to 0.5
(see text). (D) This relative log posterior represents the Old
Faithful inter-eruption intervals recorded to the nearest minute.
Notice that the discrete nature of the data is a dominant feature
as predicted by the results in the previous section.  In this
example, \texttt{optBINS} does more than choose the optimal number
of bins, it provides a warning that the data has been severely
rounded or truncated and that information has been
lost.}\label{fig:new-fig}
\end{figure}

\section{Application to Real Data}
It is important to evaluate the performance of an algorithm using
real data.  However, the greatest difficulty that such an
evaluation poses is that the correct solution is unknown at best,
and poorly-defined at worst.  As a result, we must rely on our
expectations. In these three examples, we will examine the optimal
solutions obtained using \texttt{optBINS} and compare them to the
density models obtained with both fewer and greater numbers of
bins. It is expected that the histograms with fewer number of bins
will be missing some essential characteristics of the density
function, while the histograms with a greater number of bins will
exhibit random fluctuations that appear to be unwarranted.  For
more definitive results, the reader is directed to Section
\ref{sec:demo}.

In Figure \ref{fig:new-fig}A we show the results from a data set
called `Abalone Data' retrieved from the UCI Machine Learning
Repository \cite{UCI:1998}. The data consists of abalone weights
in grams from 4177 individuals \cite{Nash:1994}.  The relative log
posterior (left) shows a flat plateau that has maximum
at $M = 14$ bins and slowly decreases thereafter.  Given this relatively flat plateau, we would
expect most bin numbers in this region to produce reasonable
histogram models. Compared to $M = 10$ bins (left) and $M = 20$
bins (right), the optimal density model with $M = 14$ bins
captures the shape of the density function best without exhibiting
what appear to be irrelevant details due to random sampling
fluctuations.

Figure \ref{fig:new-fig}B shows a second data set from the UCI
Machine Learning Repository \cite{UCI:1998} titled `Determinants
of Plasma Retinol and Beta-Carotene Levels'.  This data set
provides blood plasma retinol concentrations (in ng/ml) measured
from 315 individuals \cite{Nierenberg:1989}.  The optimal number
of bins for this data set was determined to be $M = 9$. Although a
second peak appears in the relative log posterior near $M = 15$, the
rightmost density with 15 bins exhibits small random fluctuations
suggesting that $M = 9$ is a better model.

Last, the Old Faithful data set is examined, which consists of 222
measurements of inter-eruption intervals rounded to the nearest
minute \cite{Azzalini&Bowman:1990}.  This data is used
extensively on the world wide web as an example of the
difficulties in choosing bin sizes for histograms. There exists,
in fact, a java applet developed by R. Webster West that allows
one to interactively vary the bin size and observe the results in
real time \cite{West&Ogden:1998}. In Figure \ref{fig:new-fig}D we
plot the relative log posterior for this data set.  For large
numbers of bins, the relative log posterior increases according to
(\ref{eq:picket-fence-posterior}) as described in Section
\ref{sec:dig}.  This indicates that the discrete nature of the
data (measured at a time resolution in minutes) is a more salient
feature than the overall shape of the density function. One could
have gathered more information by more carefully measuring the
eruption times on the order of seconds or perhaps tens of seconds.

It is not clear whether this missing information could affect the
results of previous studies, but we can obtain a useful density
function by adding a small uniformly distributed number to each
sample \cite{Bayman&Broadhurst} as discussed in the previous
section. Since the resolution is in minutes, we add a number
ranging from -0.5 to 0.5 minutes. The result in Figure
\ref{fig:new-fig}C is a relative log posterior that has a clear maximum at
$M = 10$ bins.  A comparison to the $M = 5$ bin case and the $M =
20$ bin case again demonstrates that the number of bins chosen by
\texttt{optBINS} results in a density function that captures the
essential details and neglects the irrelevant details.  In this
case, our method provides additional valuable information about the data set by indicating
that the discrete nature of the data was more relevant the the
underlying density function.  This implies that a sampling
strategy involving higher temporal resolution would have provided
more information about the inter-eruption intervals.

\begin{figure}
\centering
\makebox{\includegraphics[scale=0.7]{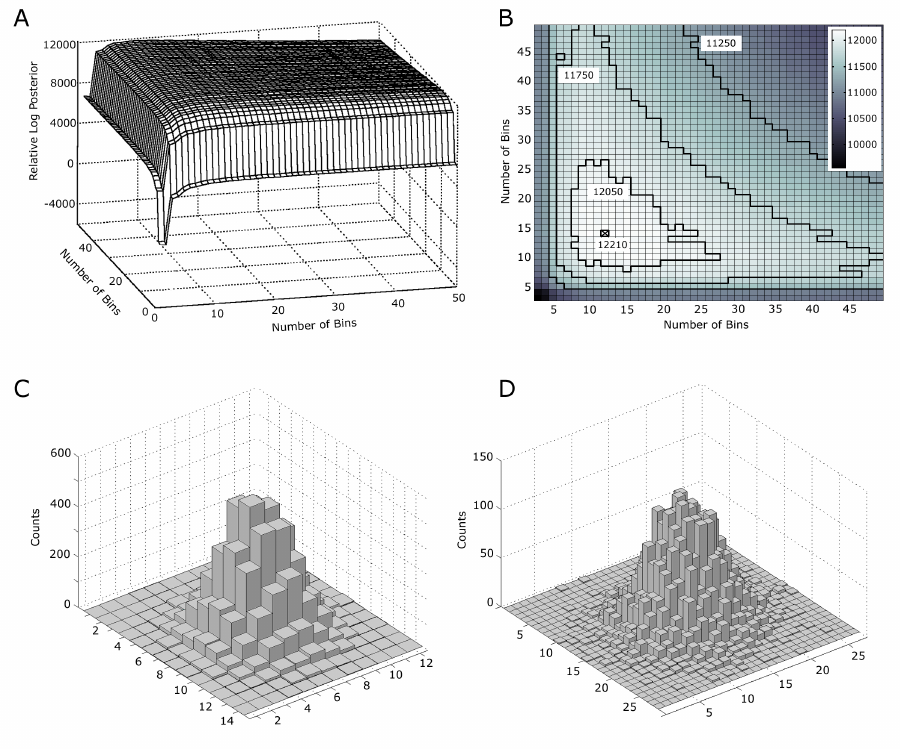}}
\caption{10000 samples were drawn from a two-dimensional Gaussian
density to demonstrate the optimization of a two-dimensional
histogram. (A) The relative logarithm of the posterior probability
is plotted as a function of the number of bins in each dimension.
The normalization constant has been neglected in this plot,
resulting in positive values of the log posterior. (B) This plot
shows the relative log posterior as a contour plot. The optimal
number of bins is found to be $12 \times 14$. (C) The optimal
histogram for this data set. (D) The histogram determined using
Stone's method has $27 \times 28$ bins. This histogram is clearly
sub-optimal since it highlights random variations that are not
representative of the density function from which the data were
sampled.}\label{fig:06}
\end{figure}

\section{Multi-Dimensional Histograms} In this section, we
demonstrate that our method can be extended naturally to
multi-dimensional histograms.  We begin by describing the method
for a two-dimensional histogram.  The constant-piecewise model
$h(x,y)$ of the two-dimensional density function $f(x,y)$ is
\begin{equation}
h(x,y;M_x,M_y) = \frac{M}{V} \sum_{j = 1}^{M_x}{\sum_{k =
1}^{M_y}{\pi_{j,k}~\Pi(x_{j-1}, x, x_j)}\Pi(y_{k-1}, y, y_k)},
\end{equation}
where $M = M_x M_y$, $V$ is the total area of the histogram, $j$
indexes the bin labels along $x$, and $k$ indexes them along $y$.
Since the $\pi_{j,k}$ all sum to unity, we have $M-1$ bin
probability density parameters as before, where $M$ is the total
number of bins. The likelihood of obtaining a datum point $d_n$
from bin $(j,k)$ is still simply
\begin{equation}
p(d_n | \pi_{j,k}, M_x, M_y, I) = \frac{M}{V}\pi_{j,k}.
\end{equation}
The previous prior assignments result in the posterior probability
\begin{equation} \label{eq:joint-posterior-multi}
p(\underline{\pi}, M_x, M_y | \underline{d}, I) \propto
\biggl(\frac{M}{V}\biggr)^N
\frac{\Gamma\bigl(\frac{M}{2}\bigr)}{\Gamma\bigl(\frac{1}{2}\bigr)^M}~
\prod_{j = 1}^{M_x}{\prod_{k =
1}^{M_y}{\pi_{j,k}^{n_{j,k}-\frac{1}{2}}}},
\end{equation}
where $\pi_{M_x,M_y}$ is $1$ minus the sum of all the other bin
probabilities.  The order of the bins in the marginalization does
not matter, which gives a result similar in form to the
one-dimensional case
\begin{equation} \label{eq:posterior-for-MxMy}
p(M_x, M_y | \underline{d}, I) \propto \biggl(\frac{M}{V}\biggr)^N
\frac{\Gamma\bigl(\frac{M}{2}\bigr)}{\Gamma\bigl(\frac{1}{2}\bigr)^M}~
\frac{\prod_{j=1}^{M_x}{\prod_{k=1}^{M_y}{\Gamma(n_{j,k}+\frac{1}{2})}}}{\Gamma(N+\frac{M}{2})},
\end{equation}
where $M = M_x M_y$.

For a D-dimensional histogram, the general result is
\begin{equation}
p(M_1, \cdots, M_D | \underline{d}, I) \propto
\biggl(\frac{M}{V}\biggr)^N
\frac{\Gamma\bigl(\frac{M}{2}\bigr)}{\Gamma\bigl(\frac{1}{2}\bigr)^M}~
\frac{\prod_{i_1=1}^{M_1}{\cdots\prod_{i_D=1}^{M_D}{\Gamma(n_{i_1,\ldots,i_D}+\frac{1}{2})}}}{\Gamma(N+\frac{M}{2})},
\end{equation}
where $M_i$ is the number of bins along the $i^{th}$ dimension,
$M$ is the total number of bins, $V$ is the $D$-dimensional volume
of the histogram, and $n_{i_1,\ldots,i_D}$ indicates the number of
counts in the bin indexed by the coordinates $(i_1, \ldots,i_D)$.
Note that the result in (\ref{eq:log-posterior-for-M}) can be used
directly for a multi-dimensional histogram simply by relabelling
the multi-dimensional bins with a single index.

Figure \ref{fig:06} demonstrates the procedure on a data set
sampled from a two-dimensional Gaussian.  In this example, 10000
samples were drawn from a two-dimensional Gaussian density. Figure
\ref{fig:06}A shows the relative logarithm of the posterior
probability plotted as a function of the number of bins in each
dimension. The same surface is displayed as contour plot in Figure
\ref{fig:06}B, where we find the optimal number of bins to be $12
\times 14$. Figure \ref{fig:06}C shows the optimal two-dimensional
histogram model. Note that the modelled density function is
displayed in terms of the number of counts rather than the
probability density, which can be easily computed using
(\ref{eq:mean-of-prob-density}) with error bars computed using
(\ref{eq:variance-of-prob-density}). In Figure \ref{fig:06}D, we
show the histogram obtained using Stone's method, which results in
an array of $27 \times 28$ bins.  
This model consists of approximately four times as many bins, and as a result, random sampling variations become visible.

\section{Algorithmic Implementations}
The basic \texttt{optBINS} algorithm takes a one-dimensional data
set and performs a brute force exhaustive search that computes the
relative log posterior for all the bin values from 1 to $M$. An
exhaustive search will be slow for large data sets that require
large numbers of bins, or multi-dimensional data sets that have
multiple bin dimensions.  In the case of one-dimensional data, the
execution time of the \texttt{Matlab} implementation (see
Appendix) on a Dell Latitude D610 laptop with an Intel Pentium M
2.13 GHz processor bilinear in both the number of data points $N$
and the number of bins $M$ to consider.  The execution time can be
estimated by the approximate formula:
\begin{equation}
T = 0.0000171 N\cdot M - 0.00026 N - 0.0026 M,
\end{equation}
where $T$ is the execution time in seconds.  For instance, for $N
= 25000$ data values and $M = 500$ bins to consider from $M = 1$
to $M = 500$, the observed time was 194 seconds, which is close to
the approximate time of 206 seconds. The algorithm is much faster
for small numbers of data points. For $N = 1000$ data points and
$M = 50$ bins, the execution time is approximately 0.5 seconds.

There are many techniques that are faster than an exhaustive
search. For example, sampling techniques, such as nested sampling
\cite{Sivia&Skilling}, are particularly efficient given that
there are a finite number of bins to consider. At this point we
have implemented the \texttt{optBINS} model and posterior in the
nested sampling framework.  The code has been designed to provide
the user flexibility in choosing the mode or the mean, which may
be more desirable in cases of extremely skewed posterior
probabilities. Such a sampling method has the distinct advantage
that the entire space is not searched. Moreover, we have added
code that stores up to 10000 log posterior results from previously
examined numbers of bins and allows us to access them later,
resulting in far fewer evaluations, especially in high-dimensional
spaces.

The computational bottleneck lies in the binning algorithm that
must bin the data in order to compute the factors for the log
posterior. We have a \texttt{Matlab} \texttt{mex} file
implementation, which is essentially compiled \texttt{C} code,
that speeds up the evaluations an order of magnitude. However, the
execution time is still constrained by this step, which depends on
both the number data points and the number of bins. The nested
sampling algorithm limits the search space by choosing the maximum
number of bins in any dimension to be $5N^{1/3}$, which on the
average is an order of magnitude greater than the number of bins
suggested by Scott's Rule (\ref{eq:scott})
Since in
one-dimension the execution time of the brute force algorithm is
bilinear in $N$ and $M$, the execution time should go as
\begin{equation}
Time \propto N\cdot M \sim M^{\frac{D+3}{3}}.
\end{equation}
We have verified this theoretical estimate with tests that show
times increasing as $M^{1.4}$ for one-dimensional data, $M^{1.6}$
for two-dimensional data, and $M^{2.0}$ for three-dimensional
data, which compare reasonably well with the predicted exponents
of $4/3$, $5/3$, and $7/3$ for one, two and three dimensions
respectively. The advantage of the nested sampling algorithm over
an exhaustive search arises from a significant reduction in the
number of calls to the binning algorithm.  This is both due to the
fact that nested sampling does not search the entire space, and
the fact that the log probabilities of previous computations are
stored for easy lookup in the event that a model with the
particular number of bins is visited multiple times.  In
one-dimension, nested sampling visits the majority of the bins and
does not significantly outperform an exhaustive search.  However
in two and three dimensions, nested sampling visits successively
fewer bin configurations, which results in remarkable speed-ups
over the exhaustive search algorithm.  For example, in
three-dimensions, exhaustive search takes 2822 seconds for 2000
data points; whereas nested sampling takes only 480 seconds.

The exhaustive search \texttt{optBINS} algorithm and the nested
sampling implementation and supporting code in \texttt{Matlab} can
be downloaded from: \url{http://knuthlab.rit.albany.edu/index.php/Products/Code}.  More recently, the \texttt{optBINS} algorithm has been coded into Python for \emph{AstroML} (\url{http://astroml.github.com/}) under the function name \texttt{knuth\textunderscore nbins} where it is referred to as \emph{Knuth's Rule} \cite{astroMLText, astroML}. \emph{AstroML} is a freely available Python repository for tools and algorithms commonly used for statistical data analysis and machine learning in astronomy and astrophysics.

\section{Discussion}
The optimal binning algorithm, \texttt{optBINS}, presented in this paper relies on finding the mode of the
marginal posterior probability of the number of bins in a
piecewise-constant density function model of the distribution from which the data were sampled. This posterior probability
originates as a product of the likelihood of the density
parameters given the data and the prior probability of those same
parameter values.  As the number of bins increases the prior probability (\ref{eq:prior-for-pi's}), which depends on the inverse of the square root of the product of the bin probabilities tends to increase.  Meanwhile, the joint likelihood (\ref{eq:likelihood}), which is a product of the bin probabilities of the individual data tends to decrease\footnote{This is the reverse from what one usually expects where increasing the number of parameters decreases the prior and increases the likelihood.}.
Since the posterior is a product of these two
functions, the maximum of the posterior probability occurs at a
point where these two opposing factors are balanced.  This
interplay between the likelihood and the prior probability
effectively implements Occam's razor by selecting the most simple
model that best describes the data.

We have studied this algorithm's ability to model the underlying density by comparing its behavior to several other popular bin selection techniques: Akaike model selection criterion \cite{Akaike:1974}, Stone's Rule \cite{Stone:1984}, and Scott's rule \cite{Scott:1979, Scott:1992}, which is similar to the rule proposed by Freedman and Diaconis \cite{Freedman&Diaconis:1981}. The proposed algorithm was found to often suggest fewer bins than the other techniques.  Given that the integrated square error between the modeled density and the underlying density tends to increase with the number of bins, one would do best to simply chose a large number of bins to estimate the density function from which the data were sampled.  This is not the goal here.  The proposed algorithm is designed to optimally describe the data in hand, and this is done by maximizing the likelihood with a noninformative prior.


The utility of this algorithm was also demonstrated by applying it
to three real data sets.  In two of the three cases the algorithm recommended reasonable bin numbers.  In the third
case involving the Old Faithful data set it
revealed that the data were excessively rounded.  That is, the
discrete nature of the data was a more salient feature than the
shape of the underlying density function.  To obtain a reasonable number of bins in this case, one need only add sufficiently small random numbers to the original data points.  However, in the example of the Old Faithful data set \texttt{optBINS} indicates something more serious.  Excessive rounding has resulted in data of poor quality and this may have had an impact on previous studies.  The fact that the \texttt{optBINS} algorithm can identify data sets where the data have been excessively rounded may be of benefit in identifying problem data sets as well as selecting an appropriate degree of rounding in cases where economic storage or transmission are an issue \cite{Knuth:2006}.

In addition to these applications, \texttt{optBINS} already has been used to generate
histograms in several other published studies.  One study by Nir et al. \cite{Nir:2006} involved making histograms of rational numbers, which led to particularly pathological `spike and void' distributions.  In this case, there is no maximum to the log posterior.  However, adding small random numbers to the data was again shown to be an effective remedy.

Our algorithm also can be readily applied to multi-dimensional
data sets, which we demonstrated with a two-dimensional data set.
In practice, we have been applying \texttt{optBINS} to
three-dimensional data sets with comparable results.  We have also
implemented a nested sampling algorithm that enables the user to
select either the most probable number of bins (mode) or the mean
number of bins.  The nested sampling implementation displays
significant speed-up over the exhaustive search algorithm,
especially in the case of higher dimensions.

It should be noted that we are working with a piecewise-constant
model of the density function, and \emph{not} a histogram \textit{per se}. The
distinction is subtle, but important.  Given the full posterior
probability for the model parameters and a selected number of
bins, one can estimate the mean bin probabilities and their
associated standard deviations.  This is extremely useful in that
it quantifies uncertainties in the density model, which can be
used in subsequent calculations.  In this paper, we demonstrated
that with small numbers of data points the magnitude of the error
bars on the bin heights is on the order of the bin heights
themselves.  Such a situation indicates that too few data exist to
infer a density function.  This can also be determined by
examining the marginal posterior probability for the number of
bins.  In cases where there are too few data points, the posterior
will not possess a well-defined mode.  In our experiments with
Gaussian-distributed data, we found that approximately 150 data
points are needed to accurately estimate the density model when
the functional form of the density is unknown.

We have made some simplifying assumptions in this work. First, the data points themselves are assumed to have no associated uncertainties.  Second, the endpoints of the density model are defined by the extreme data values, and are
not allowed to vary during the analysis. Third, we use the marginal posterior to select the optimal number of bins and then
use this value to estimate the mean bin heights and their variance. This neglects uncertainty about the number of bins, which means that the variance in the bin heights is underestimated.

Equal width bins can be very inefficient in describing
multi-modal density functions (as in Fig. \ref{fig:01}G.)  In such
cases, variable bin-width models such as the maximum likelihood
estimation introduced by Wegman \cite{Wegman:1975}, Bayesian partitioning
\cite{Denison:2002}, Bayesian Blocks \cite{Scargle:2005}, Bayesian bin distribution inference \cite{Endres&Foldiak:2005}, Bayesian regression of piecewise constant functions \cite{Hutter:2007}, and Bayesian model determination through
techniques such as reversible jump Markov chain Monte Carlo
\cite{Green:1995} may be more appropriate options in certain
research applications.

For many applications, \texttt{optBINS} efficiently delivers histograms with a number of bins that provides an appropriate depiction of the shape of the density function given the available data while minimizing the appearance
of random fluctuations.  A Matlab implementation of the algorithm is given in the Appendix and can be downloaded from \url{http://knuthlab.rit.albany.edu/index.php/Products/Code}.  A Python implementation is available from \emph{AstroML} (\url{http://astroml.github.com/}) under the function name \texttt{knuth\textunderscore nbins} where it is referred to as \emph{Knuth's Rule} \cite{astroMLText}.

\section*{Appendix: Matlab code}
\begin{verbatim}
% optBINS finds the optimal number of bins for a one-dimensional
% data set using the posterior probability for the number of bins
% This algorithm uses a brute-force search trying every possible
% bin number in the given range.  This can of course be improved.
% Generalization to multidimensional data sets is straightforward.
%
% Usage:
%           optM = optBINS(data,maxM);
% Where:
%           data is a (1,N) vector of data points
%           maxM is the maximum number of bins to consider
%
% Ref: K.H. Knuth. 2012. Optimal data-based binning for histograms
% and histogram-based probability density models, Entropy.

function optM = optBINS(data,maxM)

if size(data)>2 | size(data,1)>1
    error('data dimensions must be (1,N)');
end
N = size(data,2);

% Simply loop through the different numbers of bins
% and compute the posterior probability for each.
logp = zeros(1,maxM);
for M = 1:maxM
    n = hist(data,M);  % Bin the data (equal width bins here)
    part1 = N*log(M) + gammaln(M/2) - gammaln(N+M/2);
    part2 = - M*gammaln(1/2) + sum(gammaln(n+0.5));
    logp(M) = part1 + part2;
end

[maximum, optM] = max(logp);
return;
\end{verbatim}


\section*{Acknowledgements}

The author would like to thank the NASA
Earth-Sun Systems Technology Office Applied Information Systems
Technology Program and the NASA Applied Information Systems
Research Program for their support through the grants NASA AISRP
NNH05ZDA001N and NASA ESTC NNX07AD97A and NNX07AN04G. The author
is grateful for many valuable conversations, interactions,
comments and feedback from Amy Braverman, John Broadhurst, J. Pat Castle, Joseph Coughlan, Charles
Curry, Deniz Gencaga, Karen Huyser, Raquel Prado, Carlos Rodriguez, William Rossow, Jeffrey Scargle, Devinder Sivia, John
Skilling, Len Trejo, Jacob Vanderplas, Michael Way and Kevin Wheeler. The author is also grateful to an anonymous reviewer who caught an error in a previous version of this manuscript.
Thanks also to Anthony Gotera who
coded the histogram binning mex file and Yangxun (Billy) Chen who
helped to code the excessive rounding detection subroutine. Thanks also to the creators of AstroML (Zeljko Ivezic, Andrew Connolly, Jacob Vanderplas, and Alex Gray) who have made this algorithm freely available in the AstroML Python repository. Two
data sets used in this paper were obtained with the assistance of
UCI Repository of machine learning databases \cite{UCI:1998} and
the kind permission of Warwick Nash and Therese Stukel. The Old
Faithful data set is made available by Education Queensland and
maintained
by Rex Boggs at:
\url{http://exploringdata.net/datasets.htm}


\bibliographystyle{amsplain}
\bibliography{knuth}

\end{document}